# 130 mA/mm β-Ga$_2$O$_3$ MESFET with Low-Temperature MOVPE-Regrown Ohmic Contacts

Arkka Bhattacharyya[1], Saurav Roy[1], Praneeth Ranga[1], Daniel Shoemaker[2], Yiwen Song[2], James Spencer Lundh[2], Sukwon Choi[2], and Sriram Krishnamoorthy[1*]

[1]Department of Electrical and Computer Engineering, University of Utah, Salt Lake City, Utah, USA 84112.

[2]Department of Mechanical Engineering, The Pennsylvania State University, University Park, Pennsylvania, 16802, USA (e-mail: sukwon.choi@psu.edu).

* Corresponding author e-mail: a.bhattacharyya@utah.edu , sriram.krishnamoorthy@utah.edu

*Abstract*— We report on the first demonstration of metalorganic vapor phase epitaxy-regrown (MOVPE) ohmic contacts in an all MOVPE-grown β-Ga$_2$O$_3$ metal-semiconductor field effect transistor (MESFET). The low-temperature (600°C) heavy (n$^+$) Si-doped regrown layers exhibit extremely high conductivity with a sheet resistance of 73 Ω/sq and record low metal/n$^+$-Ga$_2$O$_3$ contact resistance of 80 mΩ.mm and specific contact resistivity of $8.3 \times 10^{-7}$ Ω.cm$^2$ were achieved. The fabricated MESFETs exhibit a maximum ON current of 130 mA/mm and a high $I_{ON}/I_{OFF}$ ratio of $>10^{10}$. Thermal characterizations were also performed to assess the device self-heating under the high current and power conditions.

β-Ga$_2$O$_3$, being an ultra-wide bandgap material (E$_g$ = 4.6 -4.9 eV), has a projected performance advantage over other predominant wide bandgap semiconductors such as SiC and GaN [1], [2]. With an added advantage of potentially being cost-effective due to the availability of large-area melt-grown bulk substrates, it has the potential to be the material of choice for next generation solid state power switching applications. Demonstration of β-Ga$_2$O$_3$-based field-effect transistors with average breakdown fields of up to 5.5 MV/cm and breakdown voltages of 8 kV not only strengthens this promise but also demonstrates the rapid progress and breakthroughs achieved in β-Ga$_2$O$_3$-based device designs and device processing technologies and high-quality material growth [3]–[9]. Further advancement of material and device engineering and processing is critical to achieving high breakdown voltage and low ON resistance with high ON current simultaneously.

Low resistance ohmic contact is an essential part of any device, apart from high mobility channel layers, to realize high performance β-$Ga_2O_3$ based transistors with high current densities and lower conduction losses. To avoid gate-recessing, various techniques such as Si-ion implantation, spin-on-glass (SOG), regrown contact layers using molecular-beam epitaxy (MBE) and pulsed-laser deposition (PLD) have been developed to realize source/drain (S/D) ohmic contacts in $Ga_2O_3$-based MOSFETs and MESFETs [10]–[14]. While regrown contacts have reportedly provided the lowest contact resistances, the $Ga_2O_3$ devices in literature which have used regrown S/D contacts have so far mostly relied on the MBE technique to regrow the heavily-doped $n^+$ regions [13]. Metalorganic vapor phase epitaxy (MOVPE) as an epitaxial growth technique has the advantage of growing β-$Ga_2O_3$ homoepitaxial films with high room-temperature electron mobility values (close to the theoretical limit) and could be promising for fabricating $Ga_2O_3$ lateral FETs with high current densities [15]–[22]. In this work, we demonstrate for the first time, using selective area epitaxy approach, the realization of low resistance regrown S/D contacts in a fully MOVPE-grown $Ga_2O_3$ lateral MESFET with high current densities and comparatively high average electric-field (without field plates or passivation).

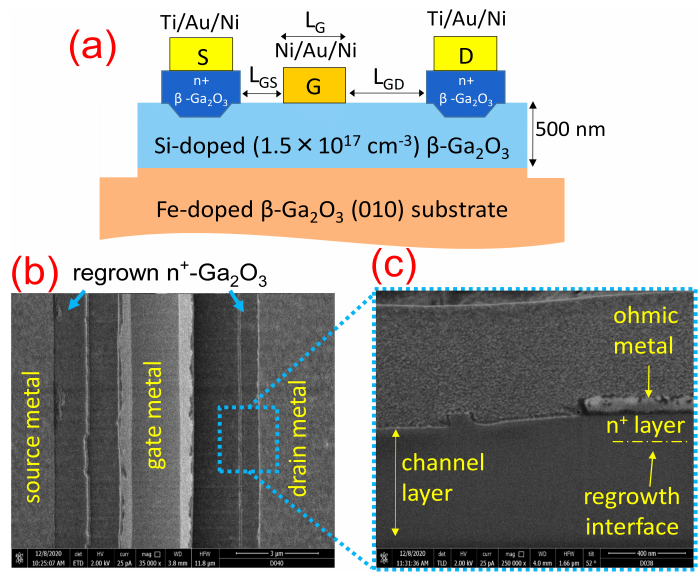

Fig 1: (a) Schematic of the fully MOVPE-grown $Ga_2O_3$ MESFET with regrown ohmic contacts. (b) Top view SEM image of the MESFET showing the regrown access regions. (c) Cross-sectional SEM of the contact region showing the estimated regrowth interface.

The epitaxial structure shown in Fig. 1(a) was grown using an Agnitron Agilis MOVPE reactor. A 500 nm thick lightly Si-doped ($1.7 \times 10^{17}$ $cm^{-3}$) β-$Ga_2O_3$ channel was grown on a (010) Fe-doped semi-insulating $Ga_2O_3$ substrate (NCT Japan) at a temperature of 810°C using triethylgallium, $O_2$ and silane gases as precursors and Ar as the carrier gas. Prior to loading into the growth reactor, the substrate was dipped in a diluted HF solution for 30 mins. From Hall measurements, the channel charge and mobility were measured to be $5.7\times10^{12}$ cm$^{-2}$ and 132 cm$^2$/Vs respectively giving a channel sheet resistance $R_{sh,ch}$ = 8.2 kΩ/□.

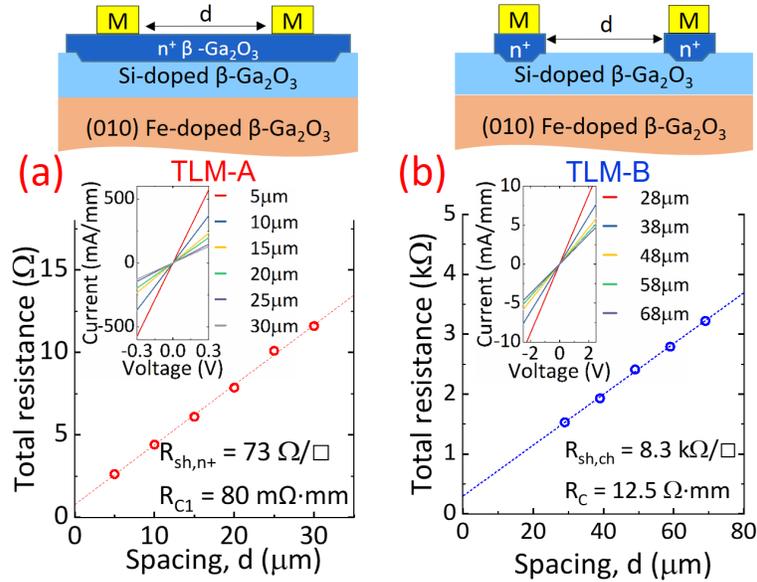

Fig 2: Schematic of TLM structure, corresponfing IV plots (inset) and total resistance vs contact spacing plots of (a) TLM pads on the MOVPE-regrown n+ Ga$_2$O$_3$ region and (b) TLM pads with patterned MOVPE-regrown n+ ohmic contacts.

The device mesa isolation was performed using a Ni/SiO$_2$ hard mask and SF$_6$/Ar plasma chemistry based ICP-RIE dry etching all the way to the substrate. A selective area MOVPE regrowth process was developed to obtain low-resistance S/D ohmic contacts. First, a sacrificial 500 nm thick SiO$_2$ layer was blanket-deposited using plasma-enhanced chemical vapor deposition. The S/D contact regions were then defined using a Ni layer patterned by photolithography and lift-off. Using Ni as the hard mask, the SiO$_2$ in the S/D contact regions was removed by the same SF$_6$/Ar plasma based directional ICP-RIE dry etching (150W RF and 600W ICP). The etching in the contact regions was extended down to the Ga$_2$O$_3$ layer with estimated Ga$_2$O$_3$ trench depths of 10-20 nm. The Ni mask was then selectively removed by dipping the sample in a diluted aqua regia solution. Next, heavily Si-doped (~$1.8\times10^{20}$ cm$^{-3}$) Ga$_2$O$_3$ was grown in the open S/D regions using MOVPE at a lowered growth temperature of 600°C. The ~100 nm thick n$^+$-Ga$_2$O$_3$ growth step was preceded by a heavy Si delta doping at the etched Ga$_2$O$_3$ surface in the contact region [15], [22]. The heavy Si delta doping was done at the etched surface to suppress any epilayer/regrown interface depletion due to any F$^-$ ion incorporation caused by the SF$_6$ dry etching. From Hall measurements on a calibration sample, the volume charge and mobility of the heavily doped n+ film were measured to be $1.4\times10^{20}$ cm$^{-2}$ and 73 cm$^2$/Vs, respectively, giving a sheet resistance of the n+ layer to be 76 Ω/□. Following the regrowth step, the polycrystalline Ga$_2$O$_3$ layer in the regions outside the contact area was removed by dissolving the sacrificial $SiO_2$ in an HF solution. Ohmic metal stack Ti/Au/Ni (20nm/100nm/30nm) was evaporated on the regrown contact regions by photolithography patterning and lift off followed by a 470°C anneal in $N_2$ for 1.5 mins. Finally, a Ni/Au/Ni (30nm/100nm/30nm) metal stack was evaporated to form the Schottky gate for the MESFET structure.

Fig.1(b) shows the top-view scanning electron microscope (SEM) image of the MESFET structure. The regrown layers were indistinguishable from the channel region and the surface morphology was similar to that of the channel. From the cross-sectional SEM (Fig.1(c)) image, no crevices or cracks were visible between the channel and the regrown region. The regrown $n^+$-$Ga_2O_3$/channel interface was very conformal with no visible interface as expected from the MOVPE growth technique.

For a uniformly doped $Ga_2O_3$ channel with $n^+$-$Ga_2O_3$ contacts, the total contact resistance ($R_C$) consists of three components: 1) $R_{C1}$ = metal/$n^+$ regrown $Ga_2O_3$ interface contact resistance, 2) $R_{n+}$ = resistance of the regrown $n^+$-$Ga_2O_3$ access region and 3) $R_{C2}$ = $n^+$ regrown $Ga_2O_3$/lightly doped $Ga_2O_3$ channel interface resistance. Each of these components were extracted from transfer length method (TLM) measurements. From TLM-A (metal contact pads spaced out on the isolated regrown $n^+$-$Ga_2O_3$ slab), sheet resistance of the regrown region ($R_{sh,n+}$) is extracted to be 73 Ω/□ which matched well with the Hall measurement done on the calibration sample mentioned earlier (Fig. 2(a)). A record low metal/$n^+$-$Ga_2O_3$ interface contact resistance ($R_{C1}$) of 80 mΩ.mm and specific contact resistance ($\rho_{C1}$) of $8.3\times10^{-7}$ Ω.cm$^2$ were achieved. This shows the low-temperature MOVPE-regrown $Ga_2O_3$ was of high quality and highly conducting. From TLM-B structure (Fig.2(b)), $R_{sh,ch}$ and total contact resistance $R_C$ are extracted to be 8.3 kΩ/□ and 12.5 Ω.mm, respectively. $R_{n+}$ is estimated from the regrown $Ga_2O_3$ access region dimensions to be 0.2 Ω.mm [23] giving $R_{C2}$ a value of 12.2 Ω.mm. This shows the total contact resistance is mainly limited by $R_{C2}$. As this is a lightly doped channel, this could be due to the etch damage at the regrowth interface caused by the F-based dry etch and could be improved by switching over to a $BCl_3$ - based dry etching. However, it is to be noted that $R_{sh,ch}$ from TLM after the regrowth process matches with that of Hall measurements done before the regrowth indicating our low-temperature regrowth process has no detrimental effect on the active region material quality.

Representative DC device output ($I_{DS}$-$V_{DS}$) and transfer ($I_{DS}$-$V_{GS}$) curves of the fully MOVPE-grown $Ga_2O_3$ MESFET are shown in Fig. 3(a) & 3(b) with device dimensions of $L_{GS}/L_G/L_{GD}$ = 1μm/1.7μm/1.6μm. The maximum normalized drain-to-source current ($I_{DS,MAX}$) was recorded to be 130mA/mm at zero gate bias and a drain bias of 15 V. The device exhibits a subthreshold slope of 122 mV/dec and a very large $I_{ON}/I_{OFF}$ ratio of $>10^{10}$ with the OFF-state current mainly limited by the measurement tool detection limit (below noise floor). From the transfer curve, a threshold voltage of -25V and a max transconductance of 4.8 mS/mm (at $V_{GS}$ = -15V) were extracted. It is to be noted that our MESFET exhibits a high ON current (130 mA/mm) without even operating at a positive $V_{GS}$ and an extremely low leakage (<10$^{-12}$ A/mm) simultaneously, thus giving our depletion-mode (D-mode) β-Ga$_2$O$_3$ channel MESFET one of the highest I$_{ON}$/I$_{OFF}$ ratio in its class. This was achieved by the proper substrate preparation to suppress the parasitic channel at the epilayer/substrate interface, maintaining a high-quality of the MOVPE-grown channel layer on the modified substrate and low-temperature device processing (low-temperature MOVPE contact regrowth) that preserved the high-mobility of carriers in the active region. Using TLM and capacitance-voltage (CV) measurements, the effective drift mobility of carriers in the channel region was estimated to be ~130 cm$^2$/Vs for V$_{GS}$=0V. Further improvement in the maximum ON currents and device performance can be achieved by realizing an accumulation channel (positive gate bias) by adopting a MOSFET structure and higher channel length to thickness ratio (higher channel aspect ratio) for improved gate control.

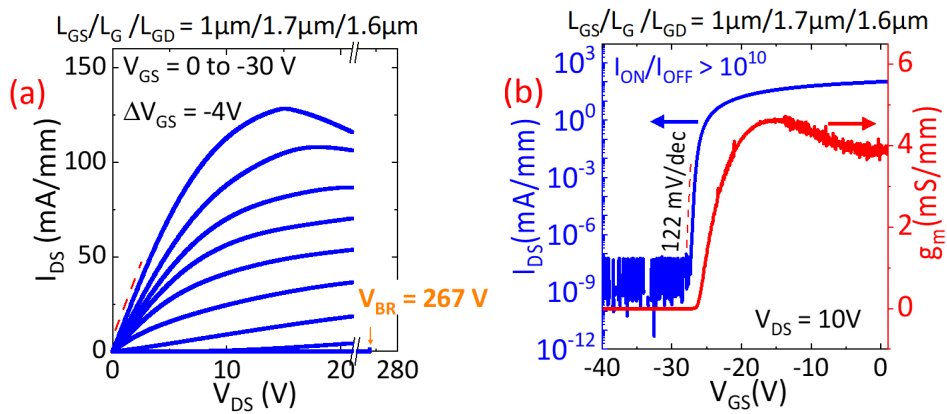

Fig 3: (a) Output characteristics and (b) Transfer characteristics of fully MOVPE-grown β-Ga$_2$O$_3$ MESFET with regrown S/D contacts.

Three-terminal OFF state breakdown measurements were performed in Fluorinert solution at V$_{GS}$ of -30V. A breakdown voltage (V$_{BR}$) of 267V was extracted for a device with L$_{GD}$ of 1.6μm. Using a 2D Sentaurus TCAD simulation of the actual structure, the peak and effective average breakdown fields were estimated to be ~7 MV/cm and 1.9 MV/cm, respectively, for this device with L$_{GD}$ = 1.6μm. With a calculated specific ON resistance (R$_{on,sp}$) of 3.7mΩ.cm$^2$, the power figure of merit is estimated to be 19 MW/cm$^2$. The R$_{on,sp}$ is the ON resistance normalized to the device active region (W×L$_{SD}$) after extracting the resistance from the linear part of the output curve at V$_{GS}$ = 0V where W and L$_{SD}$ stands for the width and the source-to-drain length of the transistor. By increasing L$_{GD}$, the maximum V$_{BR}$ measured is 778V for L$_{GD}$= 20μm. Fig. 4(b) shows the I$_{DS,MAX}$ and V$_{BR}$ as a function of L$_{GD}$ in these devices. The maximum power figure of merit (FOM) of 25MW/cm$^2$ was achieved for a device with L$_{GD}$ of 5μm. This FOM value is higher than most of the D-mode β-Ga$_2$O$_3$ channel MESFETs which do not implement any field-management or passivation techniques.

Absolute $I_{DS,MAX}$ values are measured in devices with smaller channel widths (W) while the channel length (L) was held constant (schematic in Fig.4(a) inset) that led to devices with different channel width to length ratio (W/L) across the same wafer keeping all other device dimensions ($L_{GS}/L_G/L_{SD}$ = 2.8μm/2.1μm/8.4μm) the same. For these long channel devices, the absolute $I_{DS,MAX}$ values scaled linearly with W/L as shown in Fig.4(a). The normalized $I_{DS,MAX}$ values remain constant throughout the W/L ratio. This indicates the $I_{DS,MAX}$ values in these devices are not limited by the contact resistance of the S/D ohmic contacts. In fact, the $I_{D,MAX}$ values were limited due to self-heating in these devices evident from the reduction of $I_{DS}$ value at higher $V_{DS}$ values as seen in Fig.3(a) [24], [25].

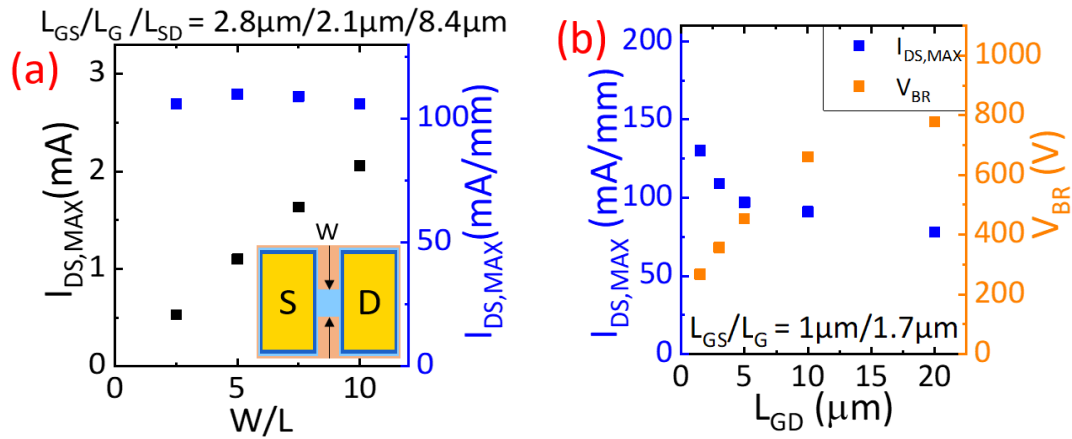

Fig 4: (a) $I_{DS,MAX}$ for devices with different channel width to length ratios (W/L) (b) $I_{DS,MAX}$ and $V_{BR}$ of devices with different $L_{GD}$.

To estimate the channel temperature at which self-heating induced current droop occurs in Fig. 3 (a) ($V_{GS}$=0 V, $V_{DS}$= 15 V, P=0.96 W/mm), the device was characterized via nanoparticle-assisted Raman thermometry using a Horiba LabRAM HR Evolution spectrometer. The temperature was measured by monitoring the $E_g$ mode frequency shift of an anatase $TiO_2$ nanoparticle deposited near the drain side corner of the gate [26]. Experimental results were validated using a three-dimensional finite element thermal model. Details of the thermal modeling procedure can be found in Ref.[26], [27]. At the current inflection point in Fig. 3(a) (P = 0.96 W/mm), the estimated channel temperature was 81°C (ΔT ~56°C). This equates to a thermal resistance of ~58.5 mm·°C/W, which is within the range of values reported in literature [28]. To visualize the device self-heating under high power conditions, infrared (IR) thermal microscopy was performed using a QFI medium wavelength infrared (MWIR) InfraScope with a 15x objective [29]. Results of the thermal analysis are illustrated in Fig. 5 (a), while an optical image of the device as well as 2D temperature plots are shown in Fig. 5 (b).

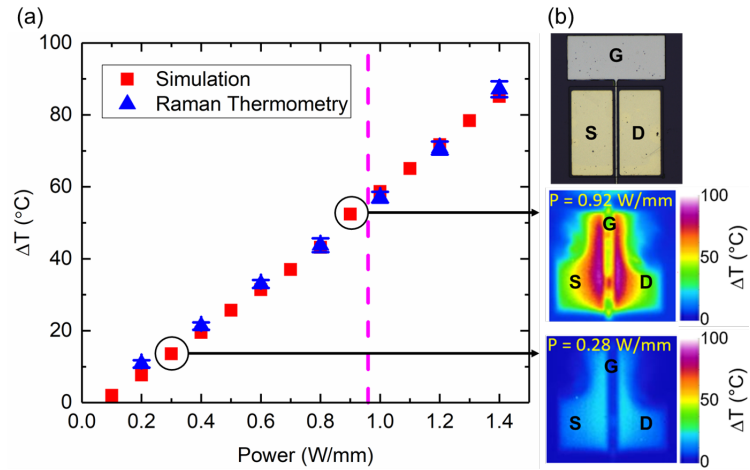

Fig 5: (a) The channel temperature rise determined by Raman thermometry and thermal modeling. The pink dashed line corresponds to the current inflection point shown in Fig. 3(a). (b) Optical image of the device under test and 2D temperature maps generated by IR thermal microscopy corresponding to power densities of 0.92 W/mm and 0.28 W/mm.

In conclusion, we demonstrate the use of MOVPE to realize low-resistance regrown ohmic contacts in a fully MOVPE-grown lateral β-$Ga_2O_3$ MESFET for the first time. The low-temperature (600°C) heavy ($n^+$) Si-doped regrown layers exhibit extremely high conductivity with sheet resistance of 73 Ω/sq and a record low metal/$n^+$-$Ga_2O_3$ contact resistance of 80 mΩ.mm and specific contact resistivity of 8.3 x $10^{-7}$ Ω.$cm^2$ were achieved. The MESFET shows a maximum current density of 130 mA/mm at zero gate bias and an extremely low leakage current (<$10^{-12}$ A/mm). Average breakdown field and maximum FOM of 1.9MV/cm and 25 MW/$cm^2$ respectively are achieved even without field-plates or passivation. However, it should be noted that the enabled high current/power capabilities must be supported by effective thermal management solutions [22]. This demonstration of first-generation fully MOVPE-grown β-$Ga_2O_3$ FETs shows the potential of MOVPE technique for realizing high ON currents in β-$Ga_2O_3$-based devices as well as a promising technique for low-temperature Ohmic contact regrowth. Further improvement in device performance can be achieved by implementing field-management techniques in a MOSFET structure along with channel engineering.

ACKNOWLEDGMENT.

This work was supported by II–VI foundation Block Gift Program. We also thank the Air Force Office of Scientific Research under Award No. FA9550-18-1-0507 (Program Manager: Dr. Ali Sayir) for financial support. Funding for Penn State was provided by the AFOSR Young Investigator Program (Grant No. FA9550-17-1-0141, Program Officers: Dr. Brett Pokines and Dr. Michael Kendra, also monitored by Dr. Kenneth Goretta) and NSF (CBET-1934482, Program Manager: Dr. Ying Sun). This